\documentclass[preprint,12pt]{elsarticle}

\usepackage[utf8]{inputenc}   
\usepackage[T1]{fontenc}
\usepackage{amsmath,amsfonts,amssymb,amsthm}

\newtheorem{theorem}{Theorem}
\usepackage{algpseudocode}
\usepackage{algorithm}
\usepackage{graphicx}
\usepackage{float}
\usepackage{hyperref}

\journal{Journal of Computational Physics}

\begin{document}

\begin{frontmatter}

\title{Nonlinear input feature reduction for data-based physical modeling}

\author[1]{Samir Beneddine}

\affiliation[1]{Aerodynamics, Aeroelasticity, Acoustic Department (DAAA), ONERA, Paris Saclay University, 8 Rue des Vertugadins, Meudon,92190, France}

\begin{abstract}
This work introduces a novel methodology to derive physical scalings for input features from data. The approach developed in this article relies on the maximization of mutual information to derive optimal nonlinear combinations of input features. These combinations are both adapted to physics-related models and interpretable (in a symbolic way). The algorithm is presented in detail, then tested on a synthetic toy model. The results show that our approach can effectively construct relevant combinations by analyzing a strongly noisy nonlinear dataset. These results are promising and may significantly help training data-driven models. Finally, the last part of the paper introduces a way to perform automatic dimensional analysis from data. The test case is a synthetic dataset inspired by the Law of the Wall from turbulent boundary layer theory. Once again, the algorithm shows that it can recover relevant nondimensional variables from data.
\end{abstract}


\end{frontmatter}


\section{Introduction}
Open physical modeling questions have recently benefited from the accelerating developments of Machine Learning tools. In particular, Deep Learning (DL) carries high hopes of tackling numerous unresolved physical problems. For instance, for the specific field of fluid mechanics, the work of \cite{singh2017machine} has been among the precursors for many papers attempting to propose new DL-augmented Reynolds-Averaged Navier-Stokes (RANS) models \cite{wu2018physics,volpiani2021machine}. To mention a few other applications, DL has also been considered as a tool to produce new Wall Models \cite{yang2019predictive}, Subgrid-Scale Models \cite{vollant2017subgrid}, and Transition Models \cite{yang2020improving}. Needless to say that this scientific trend goes well beyond the sole topic of fluid mechanics or even physics and concerns virtually all open modeling problems (for instance, in biotechnology \cite{gao2020deep}, solid mechanics \cite{haghighat2021physics}, molecular dynamics \cite{zhang2018deep}, epidemiology \cite{shorten2021deep}, etc.).

Yet, relying on a purely data-based approach has shown mitigated results. To our knowledge, no data-driven model has definitively answered any of the topics mentioned above. Indeed, neural networks (NN) are known to have poor extrapolation capabilities \cite{hettiarachchi2005extrapolation}, which results in low accuracy when a DL-model is used for physical conditions that have been unseen during training. Additionally, NNs are often black boxes from which it is hard to gain new scientific knowledge. This lack of generality and understandability of DL models may be one of the core challenges in modern Machine Learning applied to physical problems.

Consequently, recent papers have attempted to incorporate physical knowledge into these data-based approaches. Several works have directly included the governing equations of a dynamical system within the loss function of neural networks. This led for instance to the so-called Physics-Informed Neural Networks (PINN) \cite{raissi2019physics}, used for a wide range of physics-related problems \cite{cai2022physics,mao2020physics,sahli2020physics}. Another way to include physics within machine learning approaches has been done through the input features. For instance, \cite{yang2019predictive} showed in their works that scaling input features based on prior knowledge of the physics involved leads to far better results than just blindly feeding the NN with all available quantities.
Similarly, \cite{volpiani2021machine} wrote in their paper that "the choice of the input features is crucial" to obtain data-based RANS models with acceptable fidelity. These are just two examples of many works demonstrating that successful DL approaches for physical modeling depend on the judicious choice of input quantities. Therefore, the adequate definition of the input features is a central question for data-driven modeling. 

Note that feature selection is a research topic highly active in the Machine Learning community, especially on non-physics-related topics and classification tasks, with plenty of articles dedicated to the question (to name a few, \cite{dash1997feature,blum1997selection,battiti1994using,sindhwani2004feature, bollacker1996linear, tadist2019feature}).
Researchers have gathered existing methods into five main categories: filter methods, wrapper methods, embedded methods, ensemble methods, and integrative methods (see, for instance, \cite{naik2021novel, tadist2019feature, bolon2013review} for details). For this work, we are particularly interested in finding methods to identify \textit{a priori} (before any training) some relevant features just by analyzing existing data. The techniques developed in this article can be viewed as assistive tools for classical or data-driven modeling. It falls into the filter methods category, where numerous approaches have been considered based on linear correlation, Fisher score, etc. (see, for instance, \cite{naik2021novel}). 

Several particularly promising approaches for filter methods rely on information theory. Mutual information maximization (a notion explained in the paper) has been extensively used to eliminate redundant inputs among large sets of inputs. For instance, information gain is a standard tool for text classification \cite{yang1997comparative}, where mutual information between a single feature and a class
variable estimates the relevance of each feature one by one. Along the same line, \cite{battiti1994using} developed the mutual information feature selection (MIFS) algorithm for classification problems that
search greedily a set of features with high-mutual information with class labels but low mutual information among chosen features. These approaches allow for the elimination of redundant information within features. Among other works, one may cite \cite{bollacker1996linear}, which used mutual information between class variables to form optimal linear combinations of inputs. However, as mentioned by \cite{sindhwani2004feature}, these methods, which deal with continuous variables in classification tasks, demand large amounts of data and high-computational complexity.

The present work is inspired by this existing literature, except that the developed approach is tailored explicitly for physics-related problems. In contrast to the articles cited above, this study focuses on regression tasks, and we do not want to perform a simple elimination of input features. Instead, it aims to form specific nonlinear feature combinations that are both relevant for physical modeling and understandable (in a symbolic way). Additionally, the last part of the paper proposes an approach that accounts for the physical dimension of input features. The introduced algorithms are new, and they come with several implementation challenges that are addressed in the paper. Finally, one major novelty is that all methods are based on the recent work of \cite{belghazi2018mutual}, which provides innovative techniques to estimate the mutual information between two random variables. 

The paper is organized as follows. The first part defines the context of the study and illustrates the issues addressed in this article using a synthetic toy model. The complete approach and corresponding algorithm are then detailed. This algorithm is tested in the next section on the previously introduced toy model. A variant of this model is also considered to present a methodology that produces several reduced variables instead of a single one. Finally, before concluding, the last section presents a way to perform automatic dimensional analysis using all the numerical tools introduced in the paper. The method is tested on a synthetic case inspired by the Law of the Wall from the turbulent boundary layer theory. All results of the paper have been produced using the pyTorch library \cite{ketkar2021introduction}.

\section{Definitions and proposed strategy}

\subsection{Notations and definitions}
\label{sec:notations}
Let us consider a quantity $ \boldsymbol f = (f_0, f_1, \dots, f_n)\in \mathbb{R}^n$ to model. The set of input variables of the sought model is $\boldsymbol{q} = (q_0, q_1, \dots, q_m)\in \mathbb{R}^m$. In the context of fluid dynamics,  $\boldsymbol f$ may be for instance a corrective term for a RANS model (as done in \cite{parish2016paradigm} or \cite{volpiani2021machine}), and $\boldsymbol{q}$ may be some local fluid variables such as the density, velocity, molecular viscosity, shear stress tensor components, etc. 

Let us assume that $ \boldsymbol{f}$ and $\boldsymbol{q}$ are linked through an unknown stochastic model $\mathcal{M}$: $\mathcal{M}(\boldsymbol  {q}, \eta)=\boldsymbol{f}$, with $\eta$ a stochastic variable associated with some incompressible noise (that is not modeled). The initial variable set is assumed to be not optimal, i.e., it is hard to learn a data-based model $\mathcal{M}$ from $(q_0, q_1, \dots, q_m)$. It may be because this set of inputs contains redundant or useless information or because there exists a reduced set of variables that makes the mathematical model simpler (this aspect is further developed in section \ref{sec:ill_example}). Such improper input quantities will be called ''naive" input variables in the following. As mentioned in the introduction, choosing physically relevant input features is a common issue for data-driven physical modeling. More generally, the proper selection of inputs is a known problem in deep learning that may drastically affect the convergence, generality, and accuracy of the learned model (see section \ref{sec:ill_example} as an illustration of this). 

Finally, let us say that some reduced sets of more relevant input features exist. In physics, this would typically be a set of non-dimensional quantities or dimensional quantities involving an appropriate scaling. Therefore, the unknown model $\mathcal{M}$ may be split into two functions $\mathcal{E}$ and $\mathcal{M}'$, such that $\mathcal{M}=\mathcal{M}'\circ\mathcal{E}$, with $\mathcal{E}:\mathbb{R}^n\rightarrow\mathbb{R}^l$ the mapping of the naive input vector $\boldsymbol{q}$ to a better-suited $l$-dimensional space that feed the model $\mathcal{M}'$. This new model $\mathcal{M}'$ is ideally easier to learn from data and may have enhanced understandability.

\subsection{Illustrative example}
\label{sec:ill_example}
\subsubsection{Definition of the toy model}
This section presents a toy model used for a regression task using deep learning. It illustrates some intuitive ideas mentioned earlier, and it will be used later in the article to test the new approaches developed in this work.

\begin{figure}
    \centering
    \includegraphics[width=0.55\linewidth]{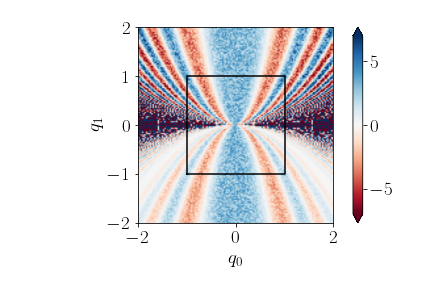}
    \caption{(Left): Toy model $f$. The black square represents the range of values for $(q_0, q_1)$ used for training neural networks to estimate $f$ ($q_1=0$ excluded).}
    \label{fig:toy_model}
\end{figure}

Let us consider a database containing 4000 triplets $(q_0^{(i)}, q_1^{(i)}$, $f^{(i)}$) obtained from the following nonlinear stochastic toy model
 \begin{equation}
    \label{eq:toymodel}
    f = \mathcal{M}(q_0, q_1, \eta) = \left(\frac{q_0^2}{q_1}+3\right)\cos(2\pi\frac{q_0^2}{q_1})(1+\eta),
 \end{equation}
with $\eta$ a white noise of amplitude 0.5, and $(q_0^{(i)}, q_1^{(i)})$ taking values in $[-1,1]\times[-1,1]\setminus \{0\}$. A graphic representation of this function is shown in figure \ref{fig:toy_model}. This toy model has been designed to be strongly nonlinear, very noisy, and such that a simple regression task by a neural network may become challenging if not done adequately. This model is particularly interesting due to its behavior near $q_1=0$: $f$ diverges and oscillates very rapidly in this neighborhood, causing difficulties to perform a regression using a neural network.

A first ``naive” strategy to predict $f$ consists of training a neural network $N$ directly from $(q_0, q_1)$ using available data. Alternatively, since $\Tilde{q} = \frac{q_0^2}{q_1}$ is a reduced variable that changes $\mathcal{M}$ into a simpler model $\mathcal{M}'$:
 \begin{equation}
    \label{eq:reducedtoymodel}
    f = \mathcal{M}'(\Tilde{q}, \eta) = \left(\Tilde{q}+3\right)\cos(2\pi\Tilde{q})(1+\eta),
 \end{equation}
one may train a network $\Tilde{N}$ to estimate $f$ from $\Tilde{q}$ using the same training data. This approach is called hereafter a "model-informed" strategy (since it requires some prior knowledge of the model). We will demonstrate some clear advantages of this latter strategy in the following. 

\subsubsection{Comparison of the naive and model-informed regression}
To perform a fair comparison between the ``naive" and the ``model-informed" strategy, $N$ and $\Tilde{N}$ have the same number of hidden layers, units per layer, and activation functions. Training is performed the same way: the database is split in two for the validation (20\% of samples) and training (80\% of samples), and an Adam optimizer \cite{kingma2014adam} is used with the same learning rate to minimize the mean square error between the network output and the actual value of~$f$. No extra penalization or more advanced technique is used here. Training is performed until an early-stopping criteria (based on the validation loss) is met. Full details are given in \ref{app:regression}.

%

\begin{figure}
    \centering
    \includegraphics[width=0.47\linewidth]{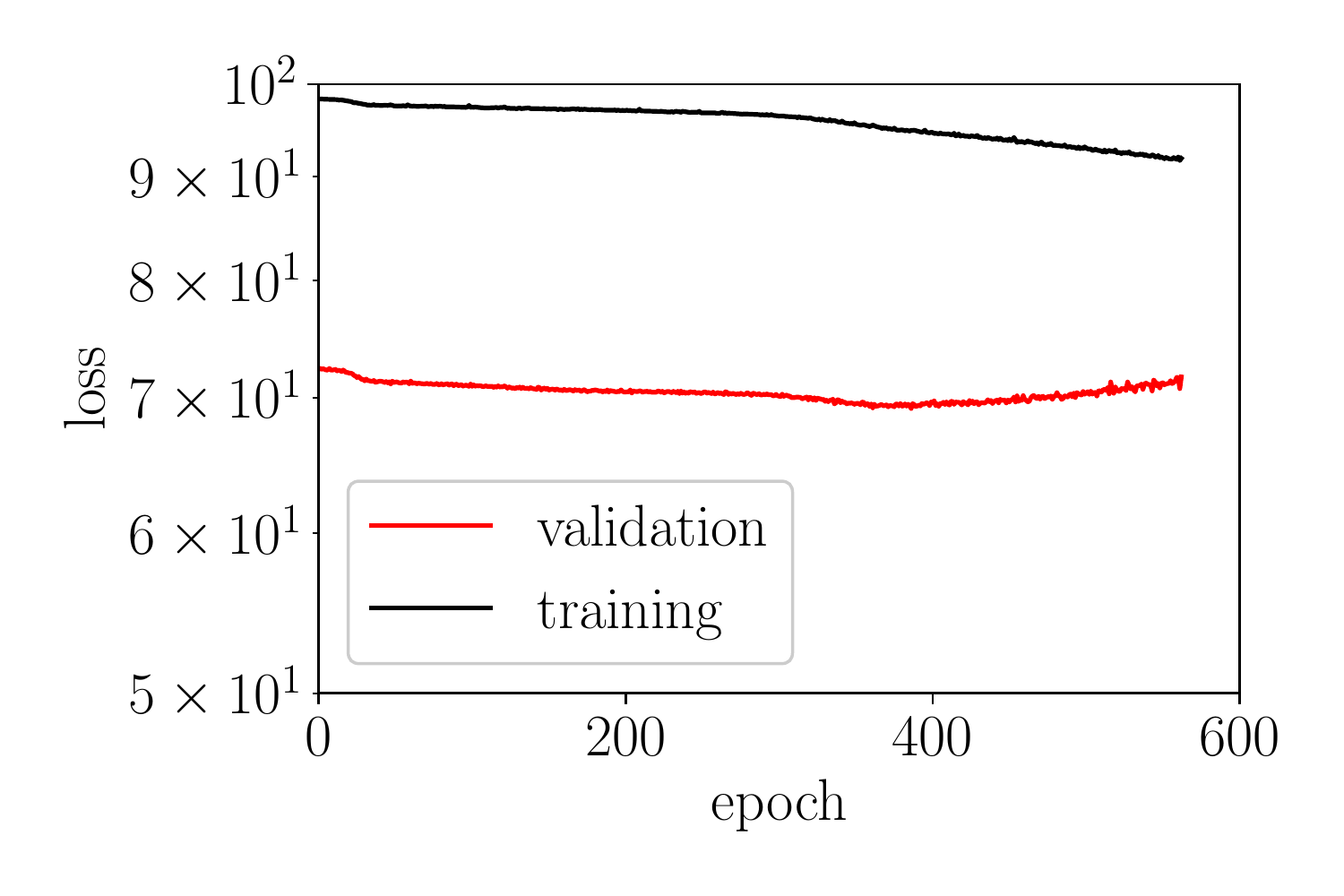}~\includegraphics[width=0.47\linewidth]{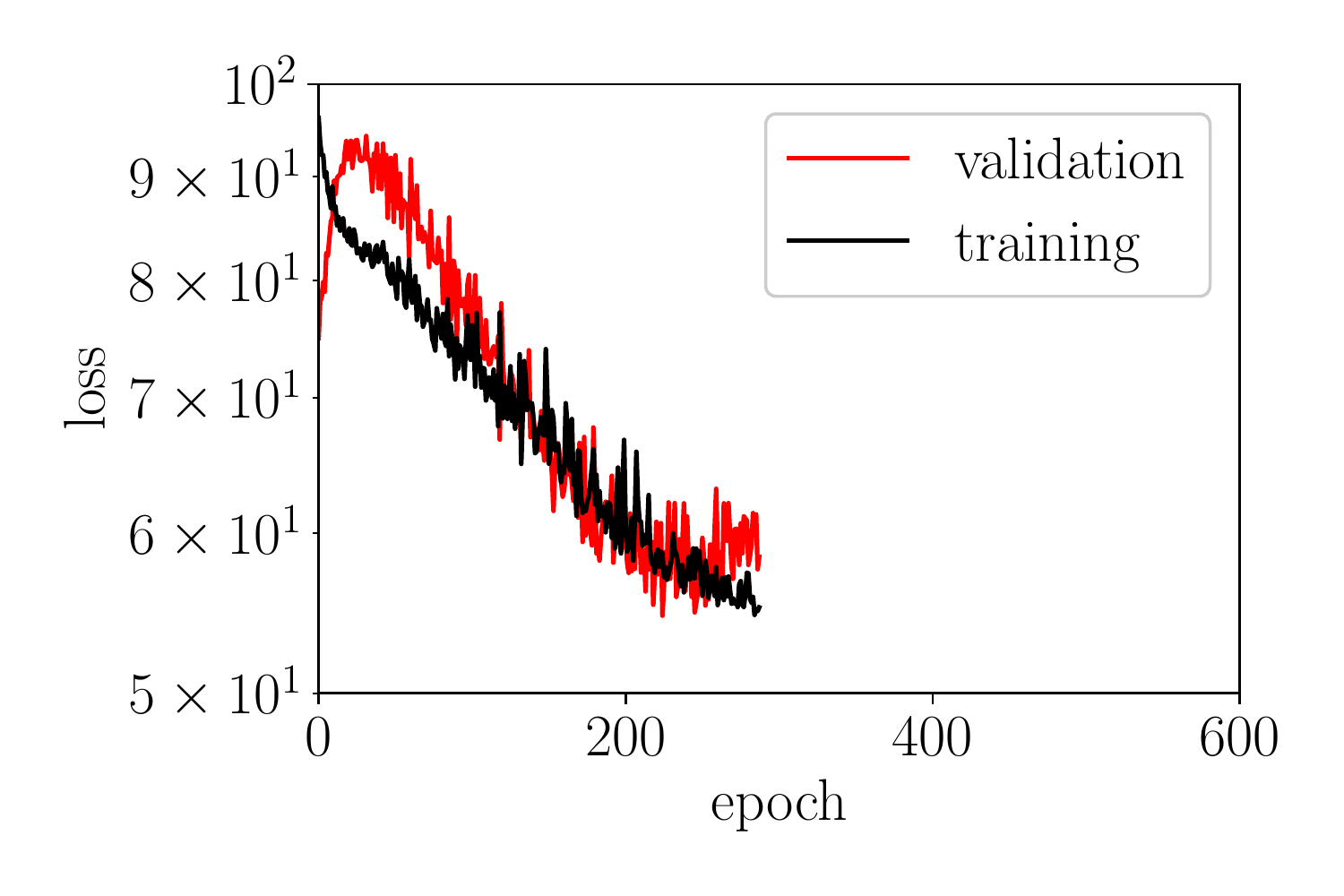}
    \caption{(Left): learning curves of $N$ (naive strategy). (Right): learning curves of $\Tilde{N}$ (model-informed strategy). The loss is the mean square error between the network output and the target value. Training is stopped based on a criterion defined in \ref{app:regression}.}
    \label{fig:learning_curves_reg_hard}
\end{figure}
\begin{figure}
    \centering
    \includegraphics[width=0.49\linewidth]{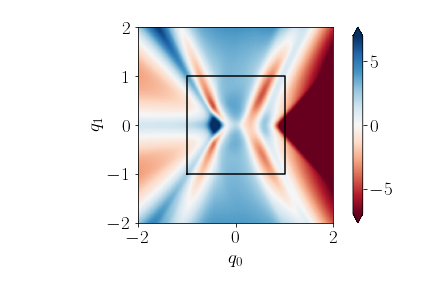}~\includegraphics[width=0.49\linewidth]{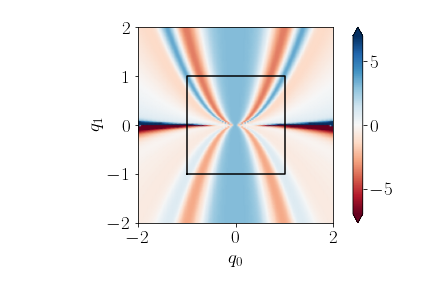}
    \caption{(Left): estimation of $f$ produced by $N$ (naive strategy). (Right): estimation of $f$ produced by $\Tilde{N}$ (model-informed strategy). The black square represents the range of values for $(q_0, q_1)$ used for training (everything outside the square shows extrapolation capabilities).}
    \label{fig:test_reg_hard}
\end{figure}

Learning curves from figure \ref{fig:learning_curves_reg_hard} show that the naive strategy has a poor and slow convergence. By comparison, the model-informed approach is much easier to train. The interpolation and extrapolation capability of each network may be seen in figure \ref{fig:test_reg_hard}. The naive approach is not only unable to extrapolate outside from the training range $(q_0^{(i)}, q_1^{(i)}) \in [-1,1]\times[-1,1]\setminus \{0\}$, but even its interpolation capabilities are unsatisfactory. On the other hand, the model-informed strategy has much better interpolation capabilities (figure \ref{fig:test_reg_hard}(right)). It is even able to extrapolate to some extent: while the center-left and center-right parts of the domain in figure \ref{fig:test_reg_hard}(right) are not well predicted, the rest shows good accuracy even for unseen values of $(q_0, q_1)$, because the corresponding values of $\Tilde{q}$ have actually been seen during training. 


These results do not demonstrate that one cannot develop a more advanced training strategy or network architecture to make the naive strategy work with this database. Still, they highlight that using specific input features may significantly ease training. This numerical experiment is a simple illustration of known neural network training behavior. Yet, it highlights well the motivations behind the present paper: the need for a strategy to deduce the "proper" reduced variable(s) to use from a database. For instance, for this toy model, it may ideally tell to use $\frac{q_0^2}{q_1}$ as an input feature. This would ease the network's training and improve its interpolation/extrapolation capabilities. 

\subsubsection{Non-unicity of reduced variables} 
The previous results show that the reduced variable $\Tilde{q}$ is a much better input feature choice than $(q_0, q_1)$. But one may wonder if other choices could be even better. In particular, for this paper, it is interesting to see if variables of the form $\Tilde{q}^x$ are also good choices (the reason for this particular form is explained in section \ref{sec:rescaling}). 

In general, as soon as $x\neq 0$, this would make a better input feature than $(q_0, q_1)$: the dimensional reduction of the input space mechanically gives a network trained with such a variable extrapolation capabilities to some extent. The only issue may be that for non-integer values of $x$, $\Tilde{q}^x$ is not be defined for negative values of $q_0$ or $q_1$, so one may favor integer values for $x$. 

\begin{figure}
    \centering
    \includegraphics[width=0.49\linewidth]{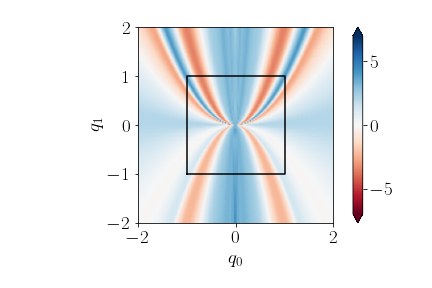}~\includegraphics[width=0.49\linewidth]{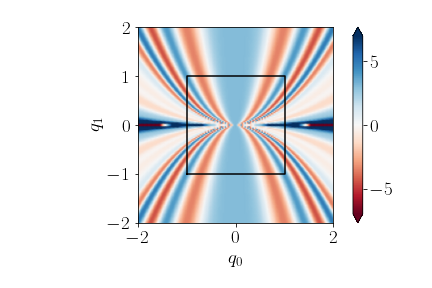}
    \caption{(Left): estimation of $f$ obtained using by $\Tilde{q}^{-1}$ as input feature. (Right): estimation of $f$ obtained using by $\Tilde{q}^{2}$ as input feature. The black square represents the range of values for $(q_0, q_1)$ used for training (everything outside the square shows extrapolation capabilities).}
    \label{fig:test_other_red}
\end{figure}

Nonetheless, most variables of the form $\Tilde{q}^x$ yield nearly equivalent results in terms of interpolation / extrapolation performances. For instance, figure \ref{fig:test_other_red}(left) shows the result field obtained from a network trained with the input feature $\Tilde{q}^{-1}$: except for the zone near $q_0=0$ (harder to learn due to the discontinuous behavior induced by this input feature choice), the overall result is similar to that from the previous section. In contrast, using  $\Tilde{q}^{2}$ provides significantly better results (figure \ref{fig:test_other_red}(right)). This is because the relation between $(q_0,q_1)$ and $f$ displays some symmetries, but the network is unaware of this and has to learn it from data. When using $\Tilde{q}^{2}$ as an input feature, the reduced model to learn $\mathcal{M}'$ has no longer any symmetry. Thus the network has less information to learn and reaches better regression capabilities. 

The conclusion will further discuss these few remarks since they provide interesting insights and maybe future work directions about the general question of choosing proper input features, which is the central topic of this article.
 
\subsubsection{Remark on correlation analysis to find relevant input features} 

One widely used tool to determine the relevancy of a given input variable to model another is the computation of correlation coefficients. The choice of this particular toy model is interesting because it shows that a correlation analysis cannot indicate that $\Tilde{q}$ is a well-suited variable: even if $\eta=0$, the Pearson correlation coefficient between $\Tilde{q}$ and $f$ for $(q_0^{(i)}, q_1^{(i)}) \in [-1,1]\times[-1,1]\setminus \{0\}$ would only be 0.04. This limitation is due to the linear nature of a correlation analysis, which would lead here to discard $\Tilde{q}$ as a relevant input variable to model $f$.

\subsection{Mutual information concept}
Before introducing a new approach for finding good input features for physical models, let us define the notion of "proper" input variable better. This paper bases the answer on information theory: a well-chosen input feature should provide as much information as possible on the quantity to model. This can be quantified using \emph{mutual information}, a concept defined below.

In information theory, the mutual information $I(X,Y)$ between two random variables $X,Y$ from a space $\mathcal{X}\times\mathcal{Y}$ is defined as

\begin{equation}
    \label{eq:mutualInf_def1}
    I(X,Y) = D_{KL}\left(P_{Y,X}\|P_X\otimes P_Y\right),
\end{equation}
where $P_{(X,Y)}$ is the joint probability distribution of the pair $(X,Y)$, $P_X$ and $P_Y$ are the marginal distribution of the pair $(X,Y)$, and $D_{KL}(\cdot\|\cdot)$ is the Kullback-Leibler (KL) divergence. After expending this expression into differential form
\begin{align}
    I(X,Y) &= \int_{\mathcal{X},\mathcal{Y}}P_X(X)P_{Y|X}(Y)\log\left(\frac{P_{Y|X}(Y)}{P_{Y}(Y)}\right)dXdY\nonumber \\
    &= -\int_{\mathcal{Y}}P_Y(X)\log P_{Y}(Y) dY\nonumber \\ +&\int_{\mathcal{X}}P_X(X)\left( \int_{\mathcal{Y}}P_{Y|X}(Y)\log P_{Y|X}(Y) dY\right)dX\nonumber\\
    &= \mathbb{E}_{Y\sim P_Y}[-\log P_Y(Y)]- \mathbb{E}_{X\sim P_X, Y\sim P_{Y|X}}[-\log P_{Y|X}(Y)],
    \label{eq:mutualInf_def2}
\end{align}
where $P_{Y|X}$ is the conditional distribution of $Y$ knowing $X$, a well-known alternative definition of $I$ involving Shannon's entropy $H$ appears
\begin{equation}
    I(X,Y) = H(Y) - H(Y|X).
    \label{eq:mutuakInf_entropy_def}
\end{equation}
Equation (\ref{eq:mutuakInf_entropy_def}) gives an alternative view on $I$: it is linked to the remaining uncertainty on $Y$ (respectively $X$) once $X$ (respectively $Y$) is known. In other words, $I(X,Y)$ is the amount of information contained in one random variable about the other. For instance, if the two variables are independent, then $H(Y|X)=H(Y)$ and $I(X,Y)=0$. Contrary to linear correlation, it is a measure of true dependence since it is able to correctly quantify nonlinear statistical dependency \cite{kinney2014equitability} (which is not the case for linear correlation as illustrated with the toy problem from section \ref{sec:ill_example}). 

The proposed strategy of this paper, detailed in the next section, relies on maximizing the mutual information between some input combinations and the quantity to model: some parameters $A$ (the weights of a neural network $\mathcal{E}_A$) will be optimized to maximize $I(\mathcal{E}_A(\boldsymbol{q}), \boldsymbol{f})$ (using notations from \ref{sec:notations}), thus providing the mapping from naive to relevant input features.

\subsection{Mutual information maximization strategy}
In most cases, it is impossible to directly maximize $I(\mathcal{E}_A(\boldsymbol{q}), \boldsymbol{f})$ because it involves the unknown posterior distribution $P_{\Tilde{\boldsymbol{q}}|\boldsymbol f}$ (using the notation $\Tilde{\boldsymbol{q}}=\mathcal{E}_A(\boldsymbol q)$). To overcome this issue, we propose an approach based on the recent work of \cite{belghazi2018mutual}, where the mutual information is first estimated using a dual representation of the KL-Divergence (providing a lower bound). This estimation is then used to maximize the mutual information.

The estimation of the mutual information relies on the Donsker-Varadhan (DV) representation \cite{donsker1983asymptotic}, based on the following theorem (see for instance \cite{belghazi2018mutual} for the proof)
\begin{theorem}
\label{th:DVrep}
Let $\Omega$ be a sample space, and $P$ and $Q$ two given probability distributions on $\Omega$. Then, the KL-divergence admits the following dual representation:
\begin{equation}
    D_{KL}(P\|Q) = \sup_{T:\Omega\rightarrow\mathbb{R}}\mathbb{E}_P[T] - \log(\mathbb{E}_Q[e^T]),
\end{equation}
with the supremum taken over all functions such that the expected values are finite.
\end{theorem}

Since equation (\ref{eq:mutualInf_def1}) defines the mutual information as the KL-divergence of the joint distribution and the product of the marginals, the DV-representation straightforwardly provides lower bounds that may be used to maximize the mutual information $I(\mathcal{E}_A(\boldsymbol{q}), \boldsymbol{f})$.

To estimate $I(X,Y)$, the idea is to consider a large family of function $T_\psi:\mathcal{X},\mathcal{Y}\rightarrow\mathbb{R}$ parametrized as a neural network with parameters $\psi\in\mathbb{R}^K$ (with $K$ the number of weights and bias of the network), and set $\psi$ such that it maximizes the quantity
\begin{equation}
    \label{eq:lower_bounde_MINE}
    \Tilde{I}_\psi(X,Y) =  \mathbb{E}_{P_{X,Y}}[T_\psi] - \log(\mathbb{E}_{P_X\otimes P_Y}[e^{T_\psi}]).
\end{equation}
This quantity is a lower bound for $I(X,Y)$, and such a network is called a \textit{statistics network} in the Mutual Information Neural Estimator (MINE) framework developed by \cite{belghazi2018mutual}. Given the universal approximation theorem for neural networks \cite{cybenko1989approximation,hornik1989multilayer}, one may expect to get through $\Tilde{I}_\psi(X,Y)$ an arbitrarily tight bound given that the network complexity is high enough. Details about the network architecture for $T_\psi$ used in the paper are given in the appendices.

\subsection{Adequate choice of function space to represent input features}
\label{sec:LN}
This section proposes a particular network architecture for $\mathcal{E}_A$ suited for physical modeling, and that provides interpretable results. When looking at existing physical models, relevant input variables come from scalings of the form \begin{equation}
    \label{eq:form_inputs}
    \Tilde{q} = \prod_i q_i^{\alpha_i}.
\end{equation}
It generally does not involve more complex functions for dimensional reasons. For instance, the logarithmic Law of the Wall, which models a part of the velocity profile of turbulent boundary layers in fluid mechanics, is a relation linking the streamwise velocity $u$ of the flow to the distance from the wall $y$ that reads \begin{equation}
\label{eq:log_law}
u^+ = \frac{1}{\kappa}\ln(y^+) + C^+,
\end{equation}
with
\begin{align*}
\tau_w = \mu \left.\frac{\partial u}{\partial y}\right|_{y=0},\\
u^* = \sqrt{\frac{\tau_w}{\rho}},\\
y^{+} = \frac{y \rho u^{*}}{\mu},\\
u^+ = \frac{u}{u^*}
\end{align*}
where $\rho$ is the fluid density, $\mu$ the dynamic viscosity, and $\kappa$ and $C^+$ two constant values. This law provides $u$ from the intermediary variables $y^+$ and $u^*$, which are combinations of the simpler variables $y, \rho, \mu, \left.\frac{\partial u}{\partial y}\right|_{y=0}$ of the form (\ref{eq:form_inputs}).

\begin{figure}[t]
    \centering
    \includegraphics[width=0.8\linewidth]{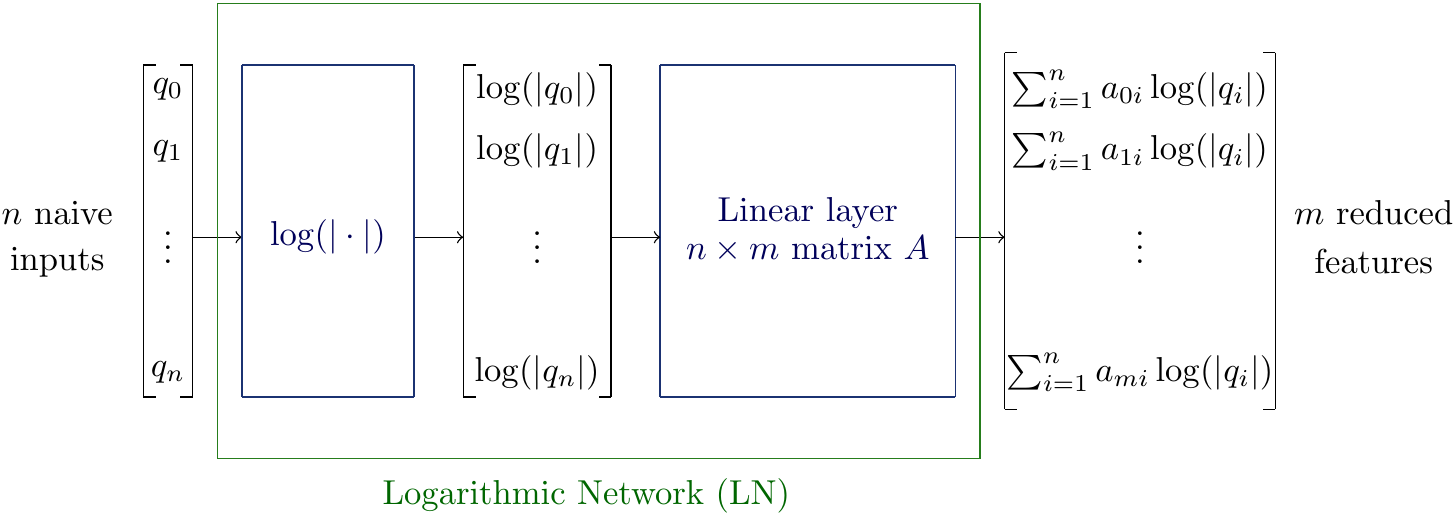}
    \caption{Logarithmic Network (LN) structure. The hyper-parameter $m$ is chosen depending on the number of input features chosen. In this article, given the strategy proposed, $m$ will always be 1.}
    \label{fig:LN}
\end{figure}

Therefore, we propose to use a logarithm representation of the quantity to model (we focus on $\log(\boldsymbol{f})$ instead of $\boldsymbol{f}$), such that the input features are searched under the form
\begin{equation}
    \label{eq:form_inputs_log}
    \Tilde{q} = \sum_i \alpha_i \log(q_i),
\end{equation}
instead of using equation (\ref{eq:form_inputs}). A neural network forming such quantities can easily be designed: it consists of a $\log$ activation followed by a linear layer with no bias. The weights of such a network are directly the exponents of equation (\ref{eq:form_inputs}). This architecture, called in the following a logarithmic network (LN), is represented in figure~\ref{fig:LN}. Note that the inputs are first made positive by taking their absolute value to handle negative numbers properly. This has no impact in the method since it is the same set of exponents $a_i$ that relates $\Tilde{q}$ with $q_i$ and $|\Tilde{q}|$ with $|q_i|$. As a side remark, an alternative solution could have been to use the complex definition of the logarithmic function
\begin{equation}
    \log(z=|z|e^{i\theta}) = log(|z|) + i\theta.
\end{equation}
One limitation remains: the LN cannot handle null values, which have to be removed or replaced by some $\epsilon$ value in the training database considered.

Note that although it is not explored in this paper, by adding a linear layer before the $\log$ activation, this approach may be straightforwardly extended to the more general form of input features
\begin{equation}
    \label{eq:form_inputs_more_complex}
    \Tilde{q} = \prod_i \left(\sum_j a_{ij} q_j\right)^{\alpha_i},
\end{equation}
which may encompass some specific cases of physical scaling that may be not covered by relation (\ref{eq:form_inputs}). In any case, different functional spaces that would be more fitted to a given problem may easily be designed and used in the framework presented in this paper.

Note that in the following, LN with a single output will be exclusively used (corresponding to $m=1$ in figure \ref{fig:LN}) due to the particular strategy used for models with multiple reduced variables (explained in section \ref{sec:mult_var}). 

\subsection{Feature Discovery algorithm}

This section combines the notion and techniques introduced above to propose the Feature Discovery (FeDis) algorithm. The idea is to maximize the mutual information between the output of an LN and a quantity to model using a MINE statistical network. The overall procedure to compute a single reduced variable is presented in algorithm \ref{alg:Fedis}. One may see that a penalization of the LN weights is added to the loss function. In the cases treated in the paper, a $L_1$ penalization was systematically considered since it promotes sparsity (and therefore helps to discard naive inputs that do not contain information about the modeled quantity) and was found to improve convergence overall. Other more advanced regularizations to handle multiple reduced variables are discussed in section \ref{sec:mult_var}. 

\begin{algorithm}
\caption{Feature Discovery (FeDis) algorithm}\label{alg:Fedis}
\begin{algorithmic}
\State  $m \gets 1$ Set the dimension of the latent space for the reduced variables to 1
\State  $A,\psi \gets$ Initialize network parameters for the 2 networks $\mathcal{E}_A$ (LN network, output dim. $=m$) and $T_\psi$
\While{loss $\mathcal{L}$  not converged} 
        \State Draw a batch of randomly sampled pair $(\boldsymbol{q}_1,\boldsymbol{f}_1),\dots,(\boldsymbol{q}_b,\boldsymbol{f}_b)$
        \State Eliminates/process pairs with $(\boldsymbol{q}_i)$ having null component(s)
        \State Form the pairs $c_1=(\mathcal{E}_A(\boldsymbol{q}_1),\log(|\boldsymbol{f}_1|)),\dots,c_b=(\mathcal{E}_A(\boldsymbol{q}_b),\log(|\boldsymbol{f}_b|))$ (joint distribution)
        \State Form ($\boldsymbol{f}'_1,\dots,\boldsymbol{f}'_b$) by shuffling ($\boldsymbol{f}_1,\dots,\boldsymbol{f}_b$) (Marginal distribution)
        \State Form the pairs $c'_1=(\mathcal{E}_A(\boldsymbol{q}_1),\log(|\boldsymbol{f}'_1|))),\dots,c'_b=(\mathcal{E}_A(\boldsymbol{q}_b),\log(|\boldsymbol{f}'_b|))$ 
        \State Evaluate the cost function $\mathcal{L} \gets - \frac{1}{b}\sum_{i=1}^b T_\psi(c_i)+\log(\frac{1}{b}\sum_{i=1}^b e^{T_\psi(c'_i)})$
        \State [Optional] Add regularization terms  $\mathcal{L} \gets \mathcal{L} + R(\theta) $
        \State Jointly update $\theta$ and $\psi$ to minimize $\mathcal{L}$ (gradient-based optimization)
\EndWhile
\end{algorithmic}
\end{algorithm}

Additionally, algorithm \ref{alg:Fedis} shows one more implementation specificity: instead of maximizing the mutual information between $\mathcal{E}(\boldsymbol{q})$ and $\boldsymbol{f}$, the algorithm uses $\mathcal{E}(\boldsymbol{q})$ and $\log(|\boldsymbol{f}|)$. This has no impact on the algorithm (maximizing the mutual information between these transformed variables still provides the input that contains the most information about $\boldsymbol f$) and has the advantage of yielding entries for $T_\psi$ of lesser magnitude (which helps avoid float-overflow issues that may occur when evaluating $e^{T_\psi(\cdot)}$).

Note that the original paper of \cite{belghazi2018mutual} about MINE networks introduced some advanced training techniques such as gradient clipping and a modified loss formulation to avoid bias during the gradient descent. These techniques have not brought any significant improvements to the results of this paper. They are therefore neither introduced nor used for the present article.

\section{Test of the FeDis algorithm}
\subsection{Results on the augmented toy model}
\label{sec:test_toy_model}
The toy model from section \ref{sec:ill_example} is used to demonstrate the ability of the FeDis approach to finding relevant reduced inputs from data generated by a noisy nonlinear model. We consider an augmented set of 14 naive input features $(q_0, q_1, \dots, q_{13})$ that add 12 useless inputs to the original dataset (to assess the ability of the approach to sort useful/useless variables). The model is therefore 
 \begin{equation}
    \label{eq:toymodel_augmented}
    f = \mathcal{M}(q_0, q_1,q_2,\dots,q_{13}, \eta) = \left(\frac{q_0^2}{q_1}+3\right)\cos(2\pi\frac{q_0^2}{q_1})(1+\eta),
 \end{equation}
with $\eta$ a white noise of amplitude 0.5. Consistently with section \ref{sec:ill_example}, the training database is made of 4000 samples $(q^{(i)}_0, \dots, q^{(i)}_{13}, f^{(i)})$, with each input $q_j$ randomly sampled in $[-1,1]\setminus \{0\}$. All details to reproduce the test case are given in the appendices. 

Figure \ref{fig:fedis_results_toy} shows the corresponding evolution of the exponent value of each input during training. As expected, all useless variables are quickly discarded, and the algorithm yields the reduced variable $q_1^{3.01}/q_0^{5.99}$ (i.e., approximately $\Tilde{q}^{-3}$). 

\begin{figure}[ht]
    \centering
    \includegraphics[height=5.0cm]{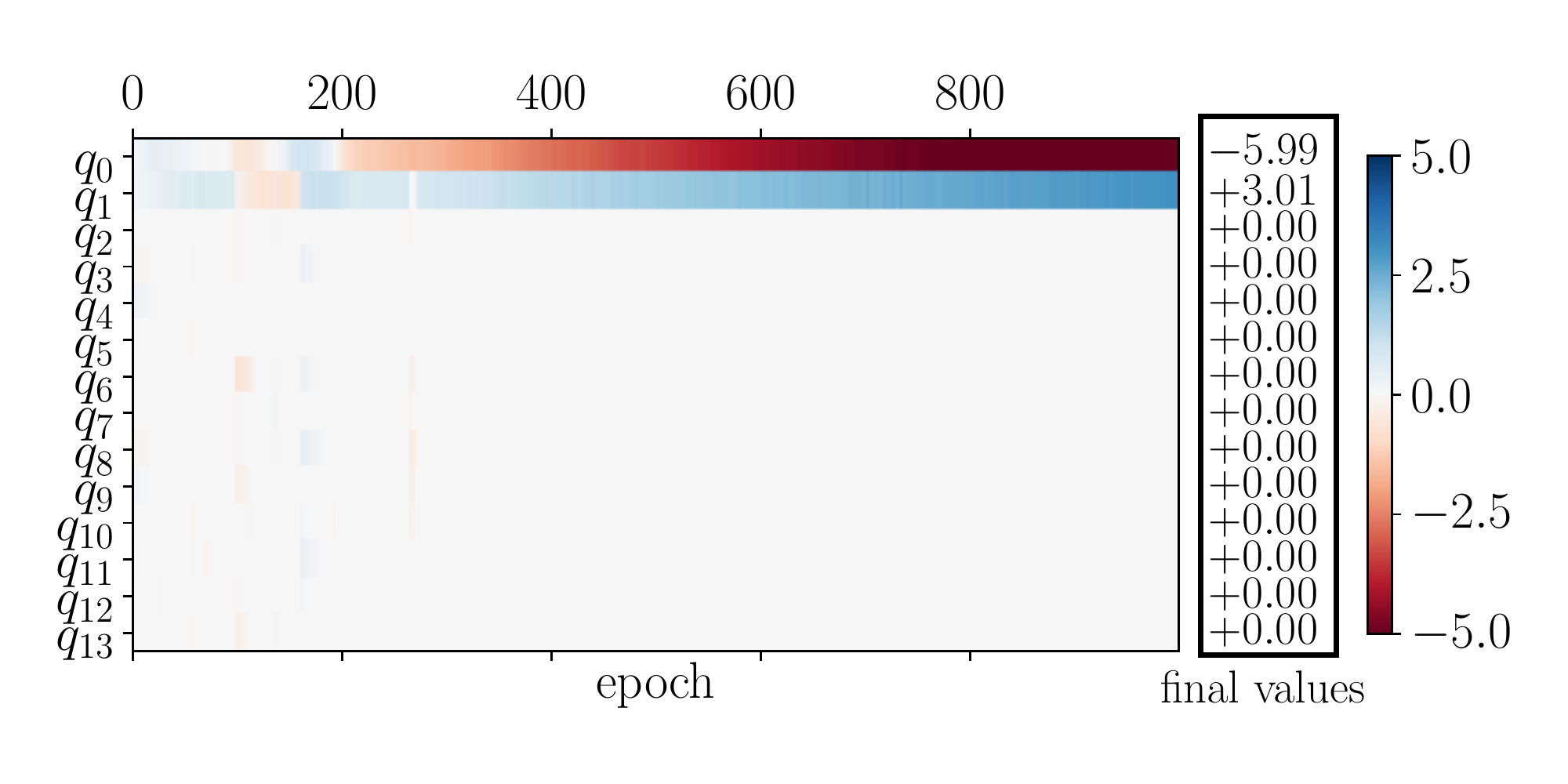}
    \caption{Evolution of the learned exponents for each input variable.}
    \label{fig:fedis_results_toy}
\end{figure}

\subsection{Normalization of the exponents}
\label{sec:rescaling}
The FeDis approach does not yield $\Tilde{q} = q_0^{2}/q_1$, but instead a variable of the form $\Tilde{q}^x$. This was expected since $I(\Tilde{q},f)=I(\Tilde{q}^x,f)$ for $x\neq0$ (the mutual information $I(X,Y)$ is preserved by any invertible deterministic transformation of $X$, see \cite{belghazi2018mutual}). Therefore, given the structure of the LN network, the maximization strategy followed here leaves the exponent $x$ undetermined (the convergence of $x$ is eventually due to the $L_1$ penalization of weights that forbid the exponents to become too large). Finding exactly $q_0^{2}/q_1$ would have been coincidental. But as mentioned in section \ref{sec:ill_example}, using $\Tilde{q}^x$ as input feature provides the same advantages as using $\Tilde{q}$. The only limitation is that non-integer exponents are not defined for negative numbers. But having this in mind, it is easy to rescale \textit{a posteriori} the exponent to overcome this issue since the method provides a symbolic representation of the input. One could also imagine penalizing non-integer exponents (for instance, using the function $p(a) = -\lambda_{int}\cos(2\pi a)$), but that would add one extra hyperparameter $\lambda_{int}$ for an uncertain and arguable improvement of the original algorithm.

A simple \textit{a posteriori} normalization to remove the undetermined nature of the exponent is given by algorithm \ref{alg:exp_rescaling}, which sets the smallest exponents to 1 (discarding nearly zero values). If applied to the results obtained in section \ref{sec:test_toy_model}, the resulting reduced variable is $\Tilde{q}^{-1} = q_1/q_0^{2}$. 

\begin{algorithm}
\caption{Exponent normalization}\label{alg:exp_rescaling}
\begin{algorithmic}
\Procedure{NormalizeExponents}{$\boldsymbol{a}=(a_0, a_1, \dots, a_{n-1})$}
  \State $a_{\max} \gets \max_i(|a_i|)$ get the highest exponent
  \State $m \gets |\boldsymbol{a}| > 0.1 a_{\max}$ build a boolean mask for values lower than 10\% of $a_{\max}$
  \State $\boldsymbol{a} \gets \boldsymbol{a} \cdot m$ set all small values to zero
  \State $\Tilde{\boldsymbol{a}} \gets nonzero(\boldsymbol{a})$ store all nonzero values
  \State $a_{\min} \gets \min_i(|\Tilde{a}_i|)$ get the smallest nonzero exponent
  \State $\boldsymbol{a} \gets \boldsymbol{a}/ a_{\min}$ normalize exponents
\EndProcedure
\end{algorithmic}
\end{algorithm}

\subsection{Dealing with multiple reduced variables}
\label{sec:mult_var}

Only one reduced variable was needed for the toy model of section \ref{sec:test_toy_model}. But more than a single reduced variable may be needed for an actual physical model. To address this question, let us consider the following modified toy model 
 \begin{equation}
    \label{eq:toymodel_augmented_2var}
    f = \mathcal{M}(q_0, q_1,q_2,\dots,q_{13}, \eta) = \left(\frac{q_0^2}{q_1}+3\right)\cos(2\pi q_0 q_1)(1+\eta).
 \end{equation}
Then, one may want to use reduced variables of the form $(\Tilde{q}_0^a,\Tilde{q}_1^b)$, with $\Tilde{q}_0=q_0 q_1$ and $\Tilde{q}_1=\frac{q_0^2}{q_1}$. A first idea would be to use an LN network with an output dimension $m=2$ (see section \ref{sec:LN}) such that the FeDis algorithm would yield a 2-dimensional vector whose components are feature combinations maximizing the information about $f$. 

In practice, this idea does not work. Indeed, the most straightforward pair of variables maximizing $I(\cdot, f)$ is the naive couple $(q_0, q_1)$ (the deterministic part of $f$ is fully defined once these two variables are known). Therefore, since the algorithm promotes sparsity, the result is likely to be $(q_0, q_1)$ (or a trivial variant). Actually, the algorithm could output nearly any pair of independent combinations of $q_0$ and $q_1$ and would still maximize the mutual information with~$f$. We need a way of promoting a combination set where each component taken individually also maximize the mutual information.

The solution is to compute one reduced variable at a time. A first run of the algorithm provides the nonlinear combination of features that maximizes $I(\cdot,f)$ (standard FeDis procedure). Then, the algorithm is rerun to give a different nonlinear combination that maximizes the mutual information, and so on. This one-by-one procedure forbids the production of trivial feature combinations such as $(q_0, q_1)$ and ensures that at each step, the next reduced variable is the best nonlinear combination given the previous ones. For this approach to work, at each step, an additional penalization term needs to be added to forbid the algorithm from producing over and over the same combination of features. 

\begin{figure}
    \centering
    \includegraphics[height=4.5cm]{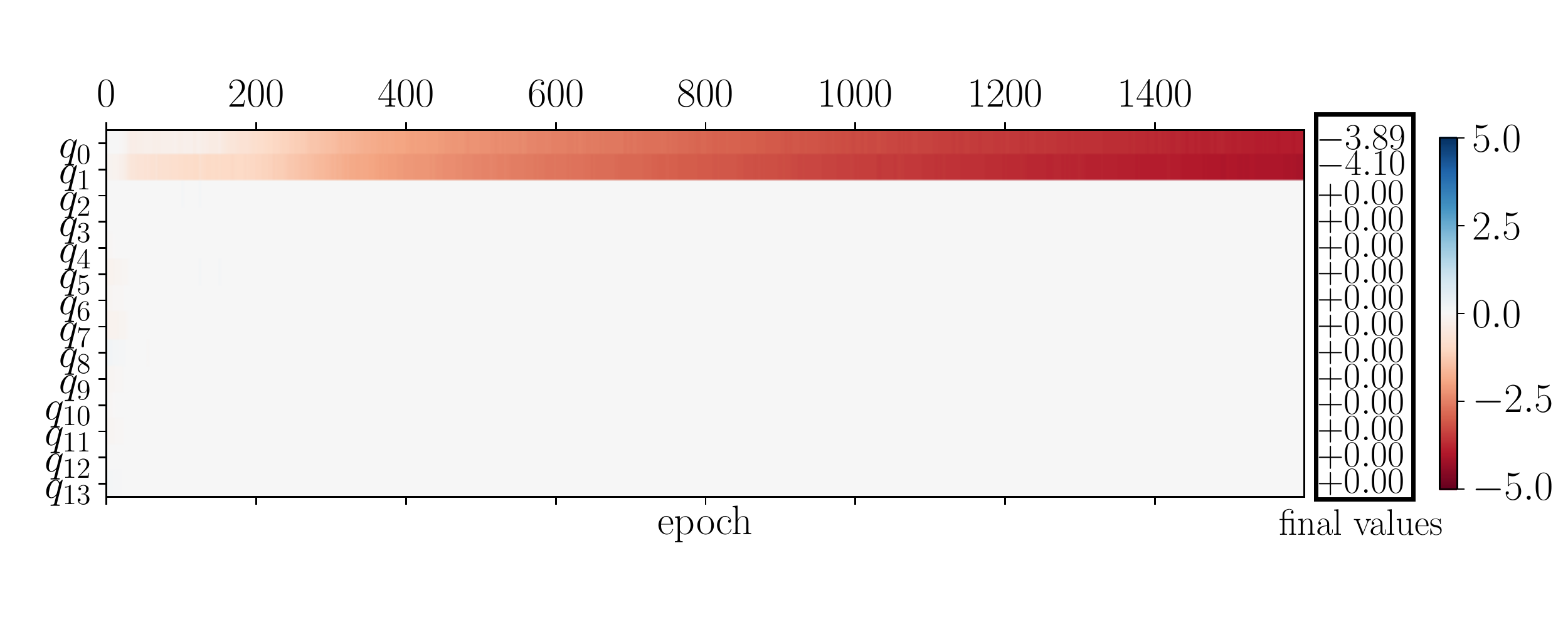} \includegraphics[height=4.5cm]{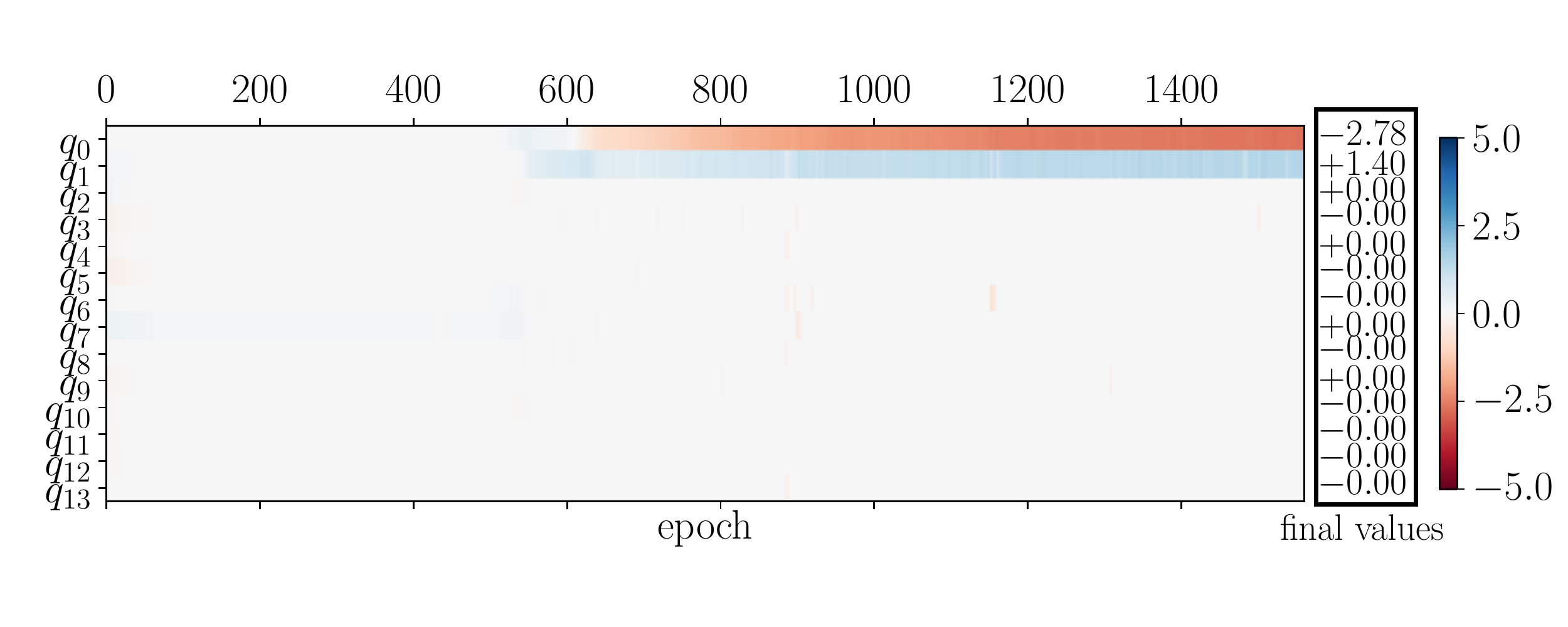}
    \caption{Evolution of the learned exponents for each input variable, second toy model (equation (\ref{eq:toymodel_augmented_2var})). (Top): first run producing the first reduced variable. (Bottom): second run producing the second reduced variable, obtained by adding a penalization promoting new input feature combinations.}
    \label{fig:fedis_2var}
\end{figure}

To illustrate the proposed approach, let us consider the second toy model defined by equation (\ref{eq:toymodel_augmented_2var}). We proceed the same way as before: the training database is made of 4000 samples $(q^{(i)}_0, \dots, q^{(i)}_{13}, f^{(i)})$, with each input $q_j$ randomly sampled in $[-1,1]\setminus \{0\}$. The database is completed by the corresponding values of $f$ generated from equation (\ref{eq:toymodel_augmented_2var}) with $\eta=0.5$. Then, reusing the same techniques as before, the FeDis algorithm is run to get a first reduced variable (details regarding the neural networks are given in \ref{app:fedis2}). The result is shown in figure \ref{fig:fedis_2var}(top). The algorithm yields the reduced variable $q_0^{-3.9} q_1^{-4.1}$, which after normalization (defined in section \ref{sec:rescaling}) gives $\Tilde{q}_0\approx q_0 q_1$ (note that if rerun and initialized differently, the network would sometimes converged toward the second reduced variable $(\frac{q_0^2}{q_1})^x$).

Then, the algorithm is rerun with a regularization promoting different feature combinations. Note that any set of exponents that would be a multiple of those already obtained are unwanted (the exponents maximizing the mutual information are defined up to a multiplicative constant).
The additional penalization term to achieve this is the following. Consider that the set of exponents obtained during the first run of the algorithm is the vector $\boldsymbol{a}=(a_0, a_1, \dots, a_n)$. Proportionality between the current exponents $\boldsymbol{A}=(A_0, A_1, \dots, A_n)$ during the rerun and $\boldsymbol{a}$ may be quantified using the normalized scalar product $s$
\begin{equation}
    s=\frac{\langle\boldsymbol{a}, \boldsymbol{A} \rangle}{\|\boldsymbol{a}\|\|\boldsymbol{A}\|}
\end{equation}
with $\langle\cdot, \cdot \rangle$ the classical dot product, and $\|\cdot\|$ the associated norm. When $\boldsymbol{a}$ and $\boldsymbol{A}$ are orthogonal, $s=0$, and when they are aligned, $s=1$. Then, the following penalization is added to the loss function of the FeDis algorithm
\begin{equation}
\label{eq:penalization_align}
    \mathcal{L}_{a} = \lambda_a e^{-\frac{(s-1)^2}{\sigma^2}},
\end{equation}
with $\lambda_a$ a weighting factor for the penalization. Note that the use of a Gaussian function instead of $s$ directly allows to not promote orthogonality of $\boldsymbol{a}$ and $\boldsymbol{A}$ (which is unwanted), but to only penalize too significant alignments. For the present study, $\sigma^2=0.2$. An exhaustive study on the optimal choice for $\sigma^2$ has not be conducted, but other values near 0.2 (0.15, 0.25) have been tested and they nearly did not changed the results. 



The results obtained using this penalization are shown in figure \ref{fig:fedis_2var}(bottom) (implementation details given in \ref{app:fedis2}). The algorithm behaves as expected, and yields the second reduced variable $q_0^{-2.8} q_1^{1.4}$, which gives after normalization $\Tilde{q}_0 = q_1/q_0^2$.

\section{Automatic dimensional analysis}
Finding relevant reduced variables to produce a physical model echoes the well-studied question of dimensional analysis. The Buckingham $\pi$-theorem states that any physically meaningful governing equation involving $k$ input variables can be reduced to an equation involving $k-l$ dimensionless parameters, with $l$ the number of physical dimensions involved. Therefore, it is natural to see if the techniques introduced in this paper can be used to automatically perform a feature reduction accounting for the physical dimension of the input variables.

\subsection{Physics-inspired test case}
The following proposes an algorithm that answers the question mentioned above. To introduce it, let us focus on the already-presented Law of the Wall,a relation between the streamwise velocity $u$ in turbulent boundary layers and the distance from the wall $y$, which links two dimensionless quantities $y^+$ and $u^+$ defined as
\begin{align}
\label{eq:yp}
y^{+} = \frac{y \sqrt{\rho} \sqrt{\left.\frac{\partial u}{\partial y}\right|_{y=0}}}{\sqrt{\mu}},\\
\label{eq:up}
u^+ = \frac{u\sqrt{\rho}}{\sqrt{\mu \left.\frac{\partial u}{\partial y}\right|_{y=0}}},
\end{align}
where $\rho$ is the fluid density, $\mu$ the dynamic viscosity, and $\kappa$ and $C^+$ two constant values. The relation, already detailed in section \ref{sec:LN}, reads
\begin{equation}
u^+ = \frac{1}{\kappa}\ln(y^+) + C^+.
\end{equation}

In the following, we consider synthetic data generated using this relation. A database made of 10000 set of values $(y, \left.\frac{\partial u}{\partial y}\right|_{y=0}, \mu, \rho, \omega_0, \omega_1, \omega_2, \omega_3)$ randomly sampled from $]0,1]^8$ has been generated. Note that this range of values has been chosen for simplicity since the case serves illustrative and demonstration purposes (thus the name of "physics-inspired" test case). It does not correspond to realistic values for each feature. Then, $u$ is evaluated for each sample using the following relation 
\begin{equation}
    u = \left(\frac{u^*}{0.41}\ln(y^+) + 5\right)(1+\eta),
\end{equation}
with $\eta$ a random white noise of amplitude 0.2. The chosen values $\kappa=0.41$ and $C^+=5$ are the standard constant values. Note that the variables $\omega_i$ are extra dummy quantities that are unused to compute $u$ (they allow to assess the ability of the approach to discard useless data).

\subsection{Description of the automatic dimensional analysis algorithm}
\begin{figure}
    \centering
    \includegraphics[width=\linewidth]{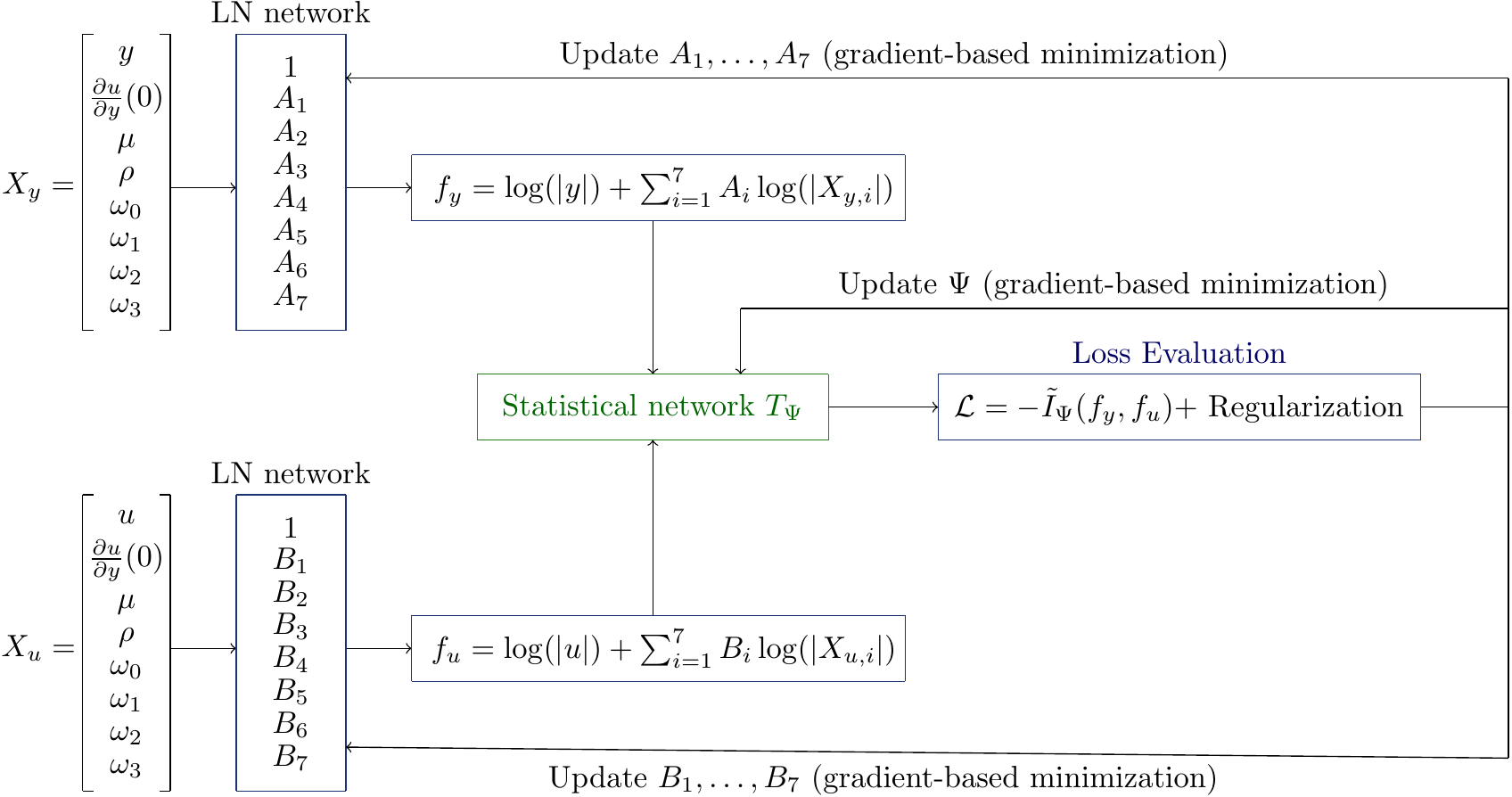} 
    \caption{Schematic representation of the FeDis variant algorithm for automatic dimensional analysis applied to the synthetic Law of the Wall case.}
    \label{fig:adim_algo}
\end{figure}
\subsubsection{Strategy}
The algorithm is described in figure \ref{fig:adim_algo}. It involves two LN networks, respectively producing a feature combination $q_y$ and $q_u$ of the form 
\begin{align}
    q_y = y\left.\frac{\partial u}{\partial y}\right|_{y=0}^{A_1} \mu^{A_2} \rho^{A_3} \omega_0^{A_4} \omega_1^{A_5} \omega_2^{A_6} \omega_3^{A_7},\\
    q_u = u\left.\frac{\partial u}{\partial y}\right|_{y=0}^{B_1} \mu^{B_2} \rho^{B_3} \omega_0^{B_4} \omega_1^{B_5} \omega_2^{B_6} \omega_3^{B_7}.
\end{align}
Note that the first weights $A_0$ and $B_0$ (the exponents of $y$ and $u$ respectively) are frozen to 1. 
Given the considered model (equations (\ref{eq:yp}) and (\ref{eq:up})), the expected outcome is $\boldsymbol{A}=(1, 0.5,-0.5,0.5,0,0,0,0)$ and $\boldsymbol{B}=(1, -0.5,-0.5,0.5,0,0,0,0)$. 
The rest of the algorithm is similar to the FeDis approach: the exponents are computed by maximizing a lower bound of the mutual information $I(q_y,q_u)$, following the same approach as that of algorithm \ref{alg:Fedis}. The specificity here comes from the penalization terms detailed in the next section.

\subsubsection{Penalization of the loss}
\label{sec:adim_pen}
In addition to the classical L1 penalization of the exponents (used for all results in the paper), two specific penalization terms have been added for this algorithm. 
The first one aims to promote non-dimensional combinations of the features. The dimensional matrices $M_{d,y}$ and $M_{d,u}$ associated with $q_y$ and $q_u$, respectively, are presented in table \ref{tab:md}. They gather the physical dimension of each feature. 
The dimension has been set arbitrarily for the dummy variables $\omega_i$. The dimension of the LN combinations corresponding to $\boldsymbol{A}$ and $\boldsymbol{B}$ is then simply given by the vector-matrix products $\boldsymbol{D}_y=\boldsymbol{A}M_{d,y}$ and $\boldsymbol{D}_u=\boldsymbol{B}M_{d,u}$ (yielding a $3\times1$ vector). The extra penalization is then based on the L2-norm of these vectors:
\begin{equation}
\label{eq:penalization_dim}
    \mathcal{L} = \lambda_d (\|\boldsymbol{D}_y\|_2^2+\|\boldsymbol{D}_u\|_2^2),
\end{equation}
with $\lambda_d$ a weighing coefficient for the penalization.

The algorithm has an easy (but unwanted) way to maximize the mutual information between $q_y$ and $q_u$. 
If the subvectors $\Tilde{\boldsymbol{A}}=(A_1, \dots, A_7)$ and $\Tilde{\boldsymbol{B}}=(B_1, \dots, B_7)$ become aligned and their components becomes really large, it may lead to $q_y\approx q_u^\alpha$, yielding  a very high mutual information $I(q_y,q_u)$. 
This is unwanted because it overlooks the relation between $u$ and $y$. 
Therefore, the alignment of $\Tilde{\boldsymbol{A}}$ and $\Tilde{\boldsymbol{B}}$ needs to be penalized. 
This is done using the same alignment penalization from section \ref{sec:mult_var} (equation (\ref{eq:penalization_align})).

\begin{table}
\centering
 \begin{tabular}{|r|ccc|} 
 \hline
   & Length $L$ & Time $T$ & Mass $M$ \\ 
 \hline
 $y$ & 1 & 0 & 0 \\ 
 $\left.\frac{\partial u}{\partial y}\right|_{0}$ & 0 & -1 & 0 \\
 $\mu$ & -1 & -1 & 1 \\
 $\rho$ & -3 & 0 & 1 \\
 $\omega_0$ & 1 & 0 & 0 \\ 
 $\omega_1$ & 0 & 0 & 1 \\ 
 $\omega_2$ & 0 & -1 & 0 \\ 
 $\omega_3$ & 1 & 1 & 1 \\ 
 \hline
 \end{tabular}~\begin{tabular}{|r|ccc|} 
 \hline
   & Length $L$ & Time $T$ & Mass $M$ \\ 
 \hline
 $u$ & 1 & -1 & 0 \\ 
 $\left.\frac{\partial u}{\partial y}\right|_{0}$ & 0 & -1 & 0 \\
 $\mu$ & -1 & -1 & 1 \\
 $\rho$ & -3 & 0 & 1 \\
 $\omega_0$ & 1 & 0 & 0 \\ 
 $\omega_1$ & 0 & 0 & 1 \\ 
 $\omega_2$ & 0 & -1 & 0 \\ 
 $\omega_3$ & 1 & 1 & 1 \\ 
 \hline
 \end{tabular}
 \caption{Dimension of the features defining $q_y$ and $q_u$. These tables define the so-called dimension matrices $M_{d,y}$ (left) and $M_{d,u}$ (right) associated with $q_y$ and $q_u$ respectively. They are used to promote dimensionless combinations.}
 \label{tab:md}
\end{table}

\subsection{Results}
\label{sec:adim_res}
\begin{figure}[t]
    \centering
    \includegraphics[height=4.5cm]{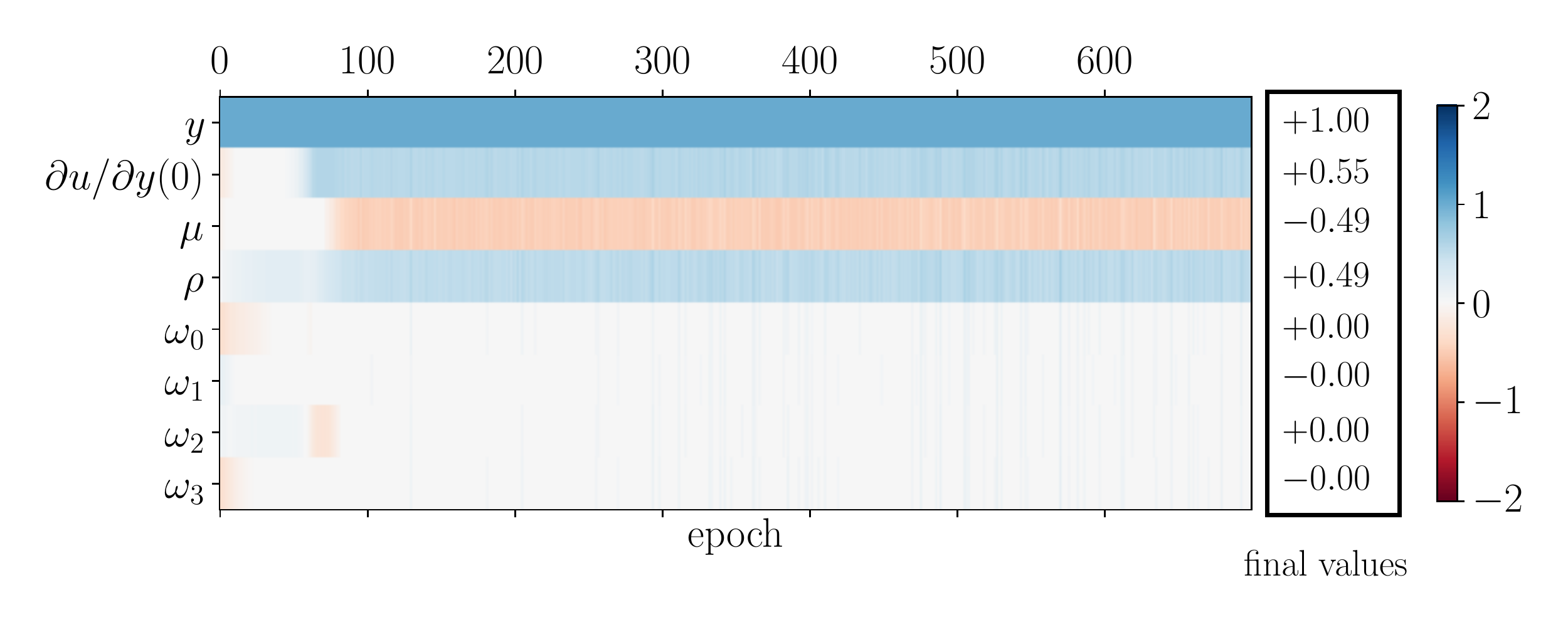} \includegraphics[height=4.5cm]{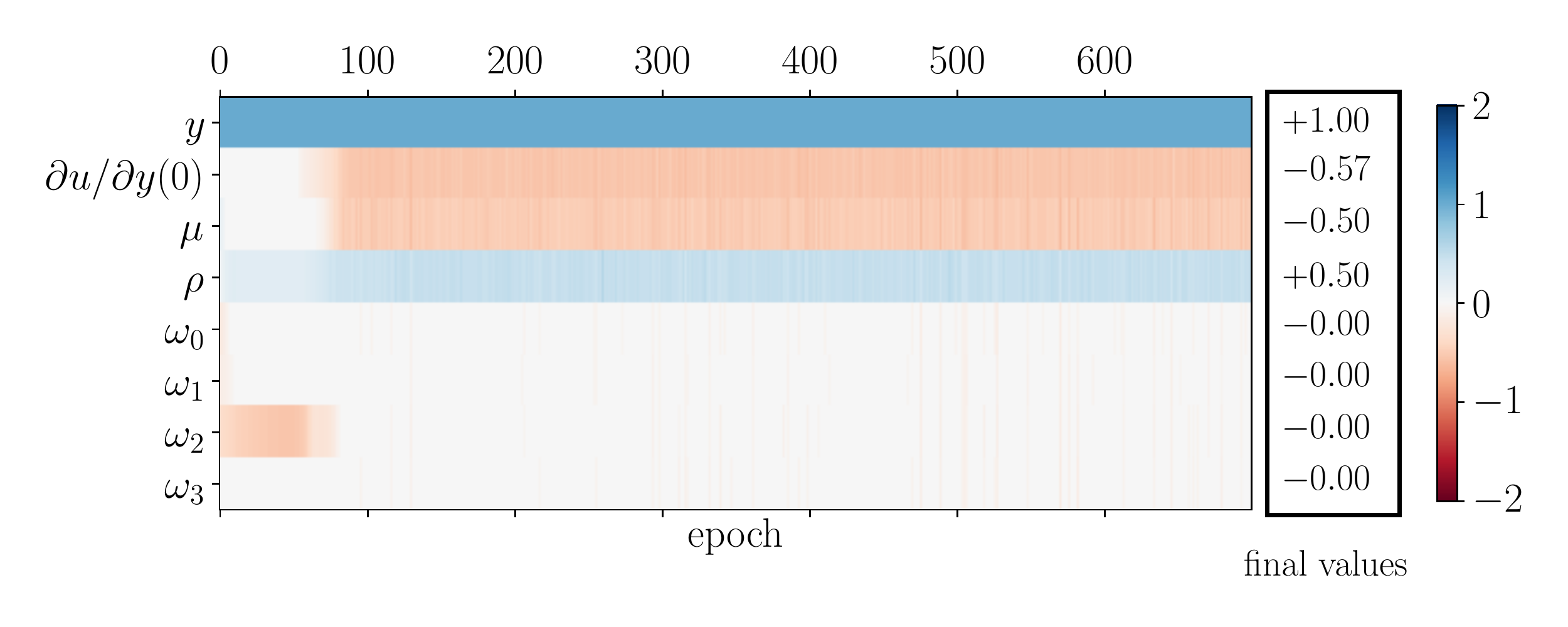}
    \caption{Evolution of the learned exponents for the Law of the Wall test case. (Top): results from the first LN network ($q_y$) (Bottom): results from the second LN network ($q_u$). The results are close to the wanted expressions defined by equations (\ref{eq:yp}) and (\ref{eq:up}).}
    \label{fig:adim_results}
\end{figure}

Details and hyperparameters used to produce the results are given in \ref{app:adim}. 
Figure \ref{fig:adim_results} shows the evolution of the exponents $\boldsymbol{A}$ and $\boldsymbol{B}$ during training. 
One may see that it produces the wanted result: the algorithm is able to nearly recover the expressions of $y^+$ and $u^+$ (equations (\ref{eq:yp}) and (\ref{eq:up})). 
One downside of the approach is that it includes multiple penalization terms. Therefore, extra additional hyperparameters need to be tuned. 
These hyperparameters have been set by trial and error. Typical behaviours when they are not set correctly is either all exponents going to zero, or $\Tilde{\boldsymbol{A}}$ and $\Tilde{\boldsymbol{B}}$ becoming proportional with a very high amplitude (see section \ref{sec:adim_pen} for the explanation).

\section{Conclusion}
This paper introduced a novel algorithm to find from a noisy dataset nonlinear input feature combinations which are optimal for physical modeling. 
They are obtained by solving a minimization problem based on the mutual information between some specific input combinations and the quantity to model. 
The algorithm has been tested on synthetic cases, showing very promising results. The last section has demonstrated that one may even use it to perform automatic dimensional analysis. 
Hopefully, this type of approach could help provide better data-driven models for physics-related problems in the near future. 
As mentioned in the article, the framework could be modified to generate different forms of feature combinations that may be better adapted to some specific modeling problem, distinct from those considered in the paper. 
Note that section \ref{sec:ill_example} has shown that some feature combinations that eliminate the symmetries of the model may provide significantly enhanced results over other reduced variables. As is, the algorithm does not account for possible symmetries in the data. 
This may be a future direction to explore to attempt to improve the algorithm.

One identified downside of the approach developed here is that it involves several hyperparameters that need to be tuned adequately (but this is an issue for most of the existing data-driven techniques, unfortunately). The paper's results were relatively robust with respect to these hyperparameters. But the synthetic cases considered were rather simple. It may be helpful to dedicate future studies to a more thorough analysis of the hyperparameter robustness in more complex cases. As the paper's primary goal was to introduce a new methodology for input feature design, exhaustive studies on the optimal choices of hyperparameters are left for future work, which may help to find empiric rules or automatic techniques to set them.

But arguably, the most interesting future work concerns the application of the FeDis technique to some of the open modeling problems mentioned in the introduction, such as RANS modeling for turbulent flows. 
The mathematical and numerical framework introduced here is general and could be applied theoretically to any dataset. Nonetheless, it is not excluded that processing actual physical data involving advanced feature dependencies may raise implementation challenges that did not appear in this first study. For instance, since the size of the synthetic datasets from the present study was small, the question of the computational cost and convergence speed has not been investigated. 
Another interesting aspect is that actual non-synthetic datasets may involve "non-homogeneous" data stemming from multiple physical phenomena. Therefore, it may sometimes be hard to get a single model for the whole dataset, and designing multiple coexisting data-driven models may be better. Each of these models may have its own distinct set of relevant features. Therefore, exploring how the present approach may be coupled with clustering techniques may be an interesting question for future work. 

\section{Acknowledgment}
This work is funded by ONERA as a part of the MODDA (MOdelization Data-Driven for Aerodynamics) project.

\appendix

\section{Regression task: section \ref{sec:ill_example}}
\label{app:regression}
The networks are made of four layers with 150 hidden units, eLU activation functions, followed by a last linear layer to produce the output. The loss function is the standard mean square error between output and target values. The network is trained using mini-batches of 100 samples. The optimizer is based on the Adam algorithm (learning rate of $10^{-4}$). Training is stopped when the validation loss exceeds the lowest value encountered so far by more than 2\% for more than thirty consecutive epochs. Then, the model's current state is dumped on disk and used for testing. This criterion is deactivated during the first fifteen epochs to avoid premature stopping. 
\section{FeDis algorithm: section \ref{sec:test_toy_model}}
\label{app:fedis1}
The function $T_\psi$ from the algorithm is a neural network made of four layers with 200 hidden units, eLU activation functions, followed by a last linear layer to produce the output (scalar value). The optimizer is a Stochastic Gradient Algorithm (SGD) with a learning rate of $7\times10^{-2}$. As mentioned in the article, an L1 penalization on the weights of the LN is added, this penalization is weighted in the total loss using a discount factor $\lambda_{L1}=5\times10^{-3}$. Training is performed using mini-batches of size 200.

\begin{figure}[ht]
    \centering
    \includegraphics[width=0.45\linewidth]{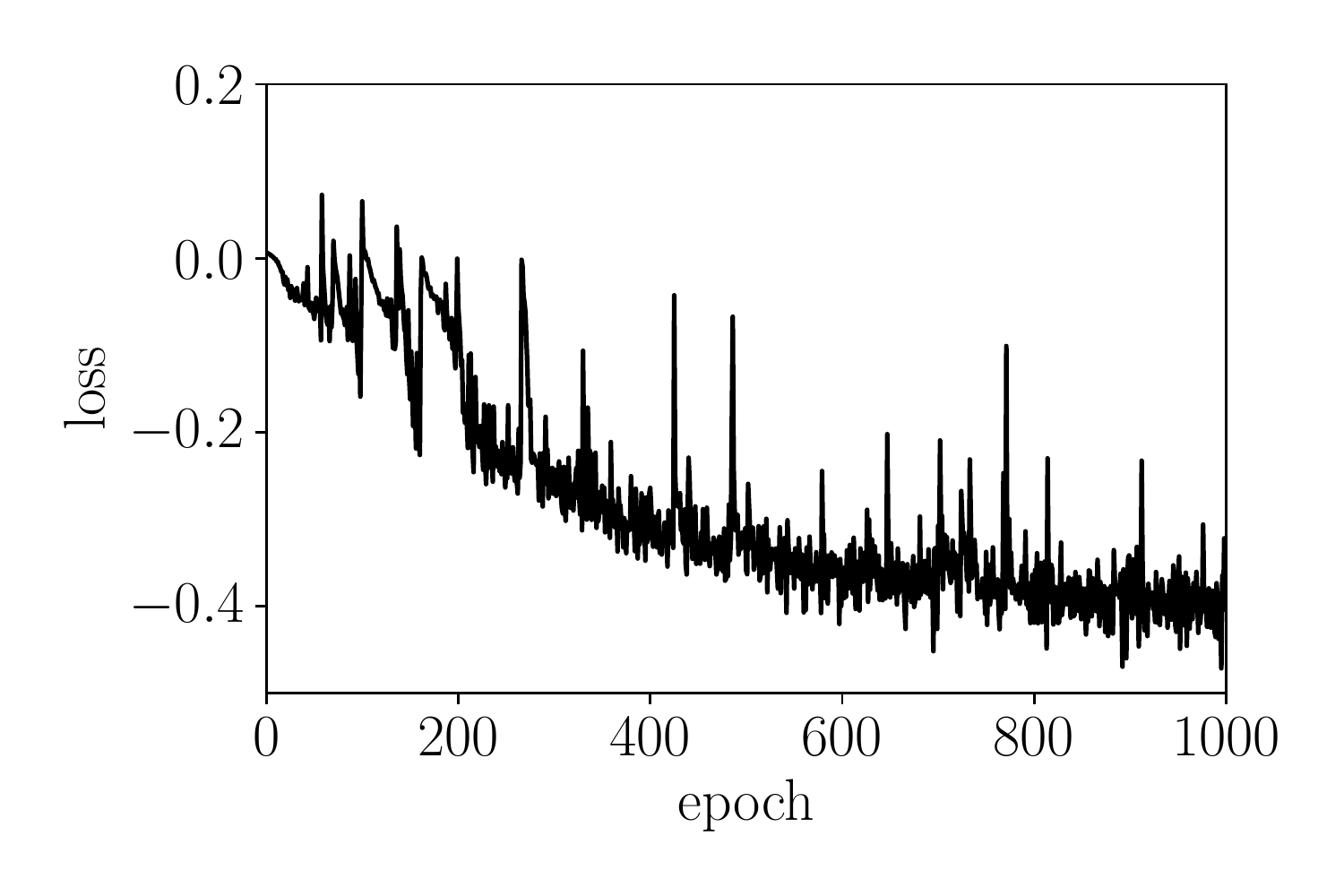}
    \caption{Loss evolution of the FeDis algorithm (toy model, equation \ref{eq:toymodel_augmented}).}
    \label{fig:fedis_loss}
\end{figure}

The losses evolution, not shown in the paper, is visible in figure \ref{fig:fedis_loss}.

\section{FeDis algorithm: section \ref{sec:mult_var}}
\label{app:fedis2}
The function $T_\psi$ from the algorithm is a neural network made of four layers with 200 hidden units, eLU activation functions, followed by a last linear layer to produce the output (scalar value). The optimizer is a Stochastic Gradient Algorithm (SGD) with a learning rate of $5\times10^{-2}$. As mentioned in the article, an L1 penalization on the weights of the LN is added, this penalization is weighted in the total loss using a factor $\lambda_{L1}=5\times10^{-3}$. Training is performed using mini-batches of size 200.

For the second run of the algorithm, the extra penalization (alignment penalization, equation (\ref{eq:penalization_align})) is weighted in the total loss using factor $\lambda_{a}=3\times10^{-2}$. All other hyperparameters are unchanged.

\begin{figure}[ht]
    \centering
    \includegraphics[width=0.45\linewidth]{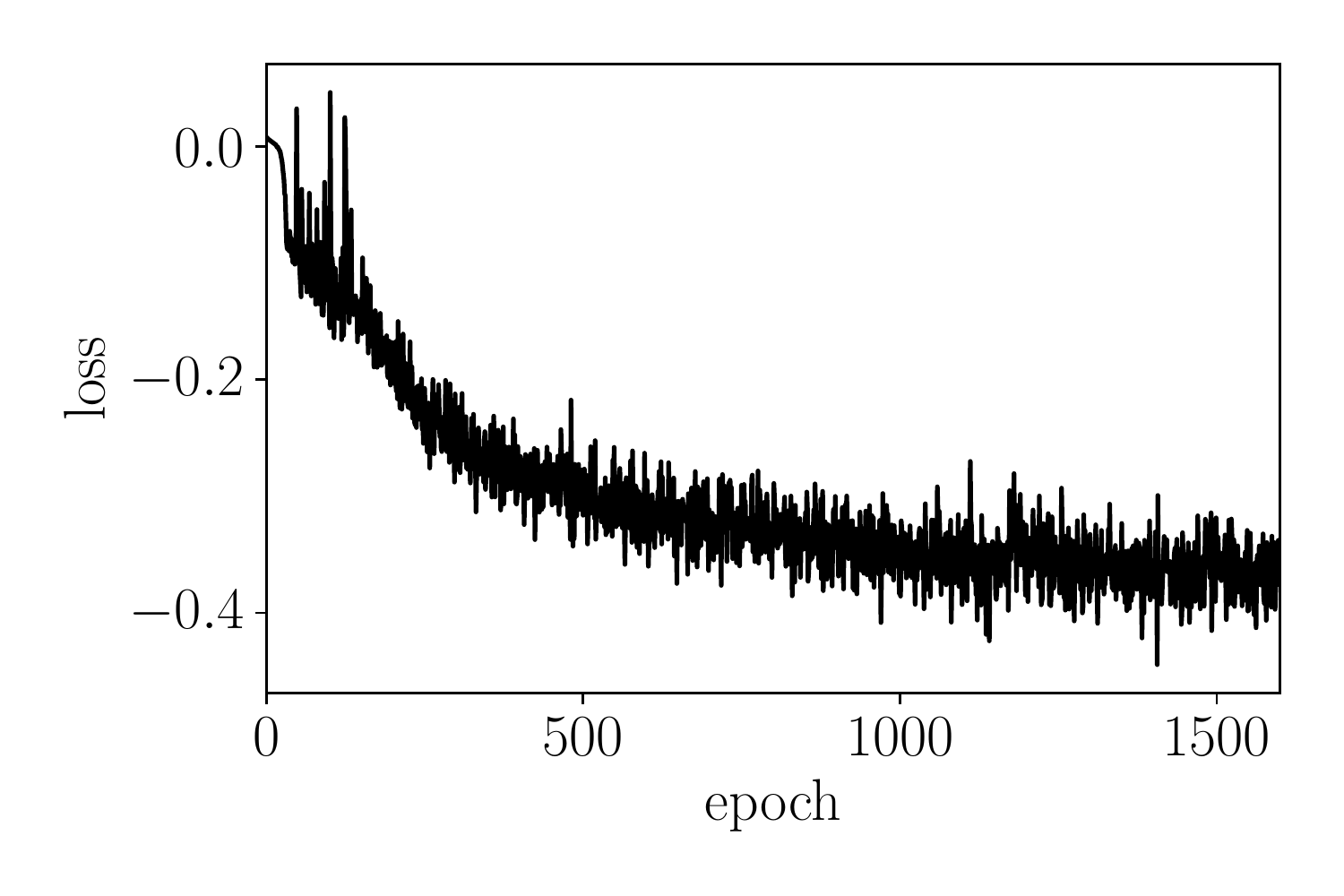}~\includegraphics[width=0.45\linewidth]{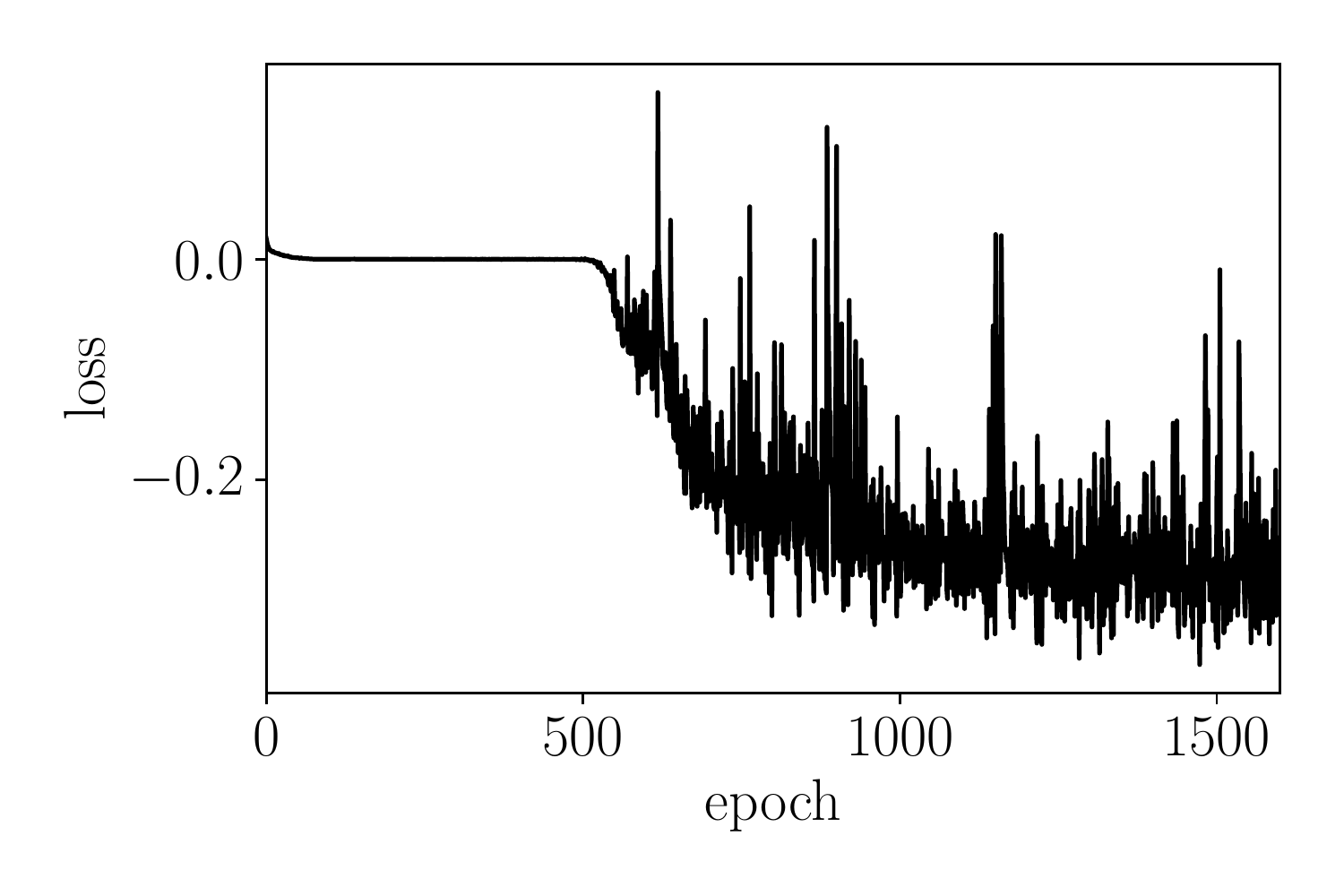}
    \caption{Evolution of the loss, second toy model (equation (\ref{eq:toymodel_augmented_2var})). (Left): first run producing the first reduced variable. (Right): second run producing the second reduced variable, obtained by adding a penalization promoting new input feature combinations.}
    \label{fig:mult_loss}
\end{figure}

The losses evolution, not shown in the paper, is visible in figure \ref{fig:mult_loss}.

\section{Automatic dimensional analysis: section \ref{sec:adim_res}}
\label{app:adim}
The function $T_\psi$ from the algorithm is a neural network made of four layers with 200 hidden units, eLU activation functions, followed by a last linear layer to produce the output (scalar value). The optimizer is a Stochastic Gradient Algorithm (SGD) with a learning rate of $2\times10^{-2}$. As mentioned in the article, an L1 penalization on the weights of each LN is added, this penalization is weighted in the total loss using a factor $\lambda_{L1}=5\times10^{-2}$. Training is performed using mini-batches of size 400.

The first extra penalization (dimensionless promotion (\ref{eq:penalization_dim})) is weighted in the total loss using factor $\lambda_{d}=5\times10^{-2}$. 
The second extra penalization (alignment penalization, equation (\ref{eq:penalization_align})) is weighted in the total loss using factor $\lambda_{a}=1$. 

\begin{figure}[ht]
    \centering
    \includegraphics[width=0.45\linewidth]{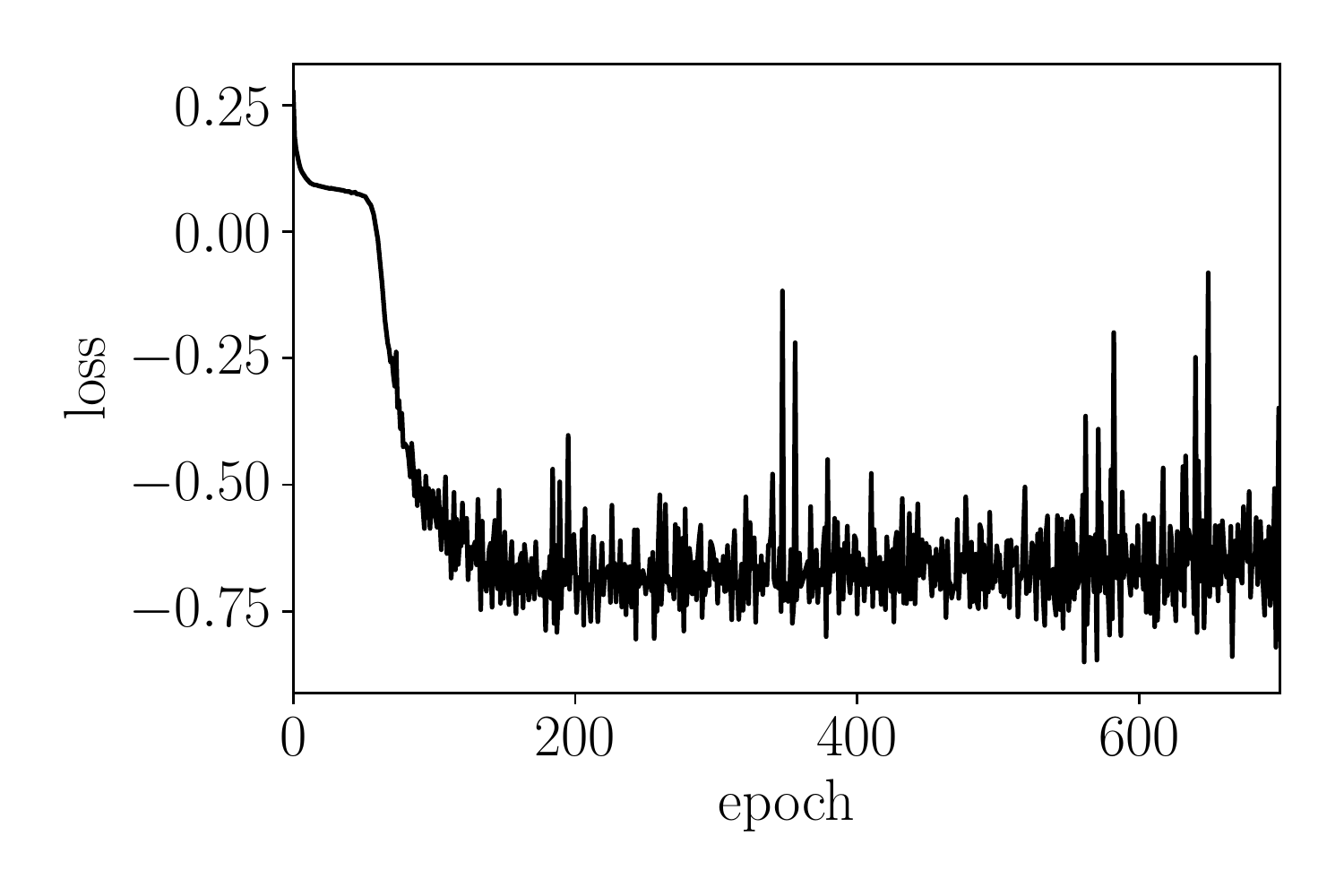}
    \caption{Evolution of the loss, synthetic Law of the Wall case.}
    \label{fig:adim_loss}
\end{figure}

The loss evolution, not shown in the paper, is visible in figure \ref{fig:adim_loss}.

\bibliographystyle{elsarticle-num} 
\bibliography{biblio}

\end{document}